# AI-driven Inverse Design System for Organic Molecules

Seiji Takeda[1,*], Toshiyuki Hama[1], Hsiang-Han Hsu[1], Toshiyuki Yamane[1], Koji Masuda[1], Victoria A. Piunova[2], Dmitry Zubarev[2], Jed Pitera[2], Daniel P. Sanders[2], and Daiju Nakano[1]

[1]IBM Research – Tokyo, Shinkawasaki, Kawasaki, Japan
[2]IBM Research – Almaden, San Jose, California, U.S.

*Corresponding author: seijitkd@jp.ibm.com

Abstract

Designing novel materials that possess desired properties is a central need across many manufacturing industries. Driven by that industrial need, a variety of algorithms and tools have been developed that combine AI (machine learning and analytics) with domain knowledge in physics, chemistry, and materials science. AI-driven materials design can be divided to mainly two stages; the first one is the modeling stage, where the goal is to build an accurate regression or classification model to predict material properties (e.g. glass transition temperature) or attributes (e.g. toxic/non-toxic). The next stage is design, where the goal is to assemble or tune material structures so that they can achieve user-demanded target property values based on a prediction model that is trained in the modeling stage. For maximum benefit, these two stages should be architected to form a coherent workflow. Today there are several emerging services and tools for AI-driven material design, however, most of them provide only partial technical components (e.g. data analyzer, regression model, structure generator, etc.), that are useful for specific purposes, but for comprehensive material design, those components need to be orchestrated appropriately. Our material design system provides an end-to-end solution to this problem, with a workflow that consists of data input, feature encoding, prediction modeling, solution search, and structure generation. The system builds a regression model to predict properties, solves an inverse problem on the trained model, and generates novel chemical structure candidates that satisfy the target properties. In this paper we will introduce the methodology of our system, and demonstrate a simple example of inverse design generating new chemical structures that satisfy targeted physical property values.

1. Introduction

Innovation in material science is a key driving force for the advancement of all manufacturing industries (i.e., those making automobiles, aircraft, solar cells, healthcare devices, pharmaceuticals, and so on). Industrial product design relies on available materials, so freedom of the product design is severely restricted by the variety of available materials. In order to release product design from restrictions due to material availability and performance, and realize more drastic product differentiation, efficient design of novel materials possessing user-demanded properties is of primary importance.

Today's material development process is well-known to stand on myriad repeats of experimental and computational trial-and-error processes that are based on the knowledge, experience, and intuition of human subject-matter-expert (SME). However, it should be noticed that the number of today's existing materials is only on the order of $10^9$ and the estimated number of theoretically configurable materials is much more than $10^{60}$ (Bohacek et al., 1996). In addition to this extra vast white space, the overwhelmingly increasing volume of documents and papers about material science have already exceeded what SME can explore. For the purpose of accelerating the speed of material development − a process that typically ranges as long as 10 to 20 years − and also boosting the freedom of material design beyond human knowledge and bias, AI has started to play a key role in materials design, often referred as "Cheminformatics" or "Materials Informatics". The advent of AI has been drastically changing the practice of material design. Various AI approaches are applied to each functional block in the general workflow of material design (see Figure 1). At the data preparation step, natural language processing (NLP) – based AI performs data extraction to generate a structured table, and also to generate a knowledge graph (KG), that is used to identify word-based "hypothesis" useful in designing materials (Choi et al., 2018). In the design process, AI-driven automated design algorithms can generate a lot of candidates rapidly with extended design freedom. At the simulation step, some kinds of physical simulation can be accelerated by shortcutting iterative steps using AI predictions. Finally, in experimental validation, the synthetic lab process is enhanced by reinforcement learning with robotics (Li et al., 2018). The focus of the present paper is the AI-driven automated design of materials.

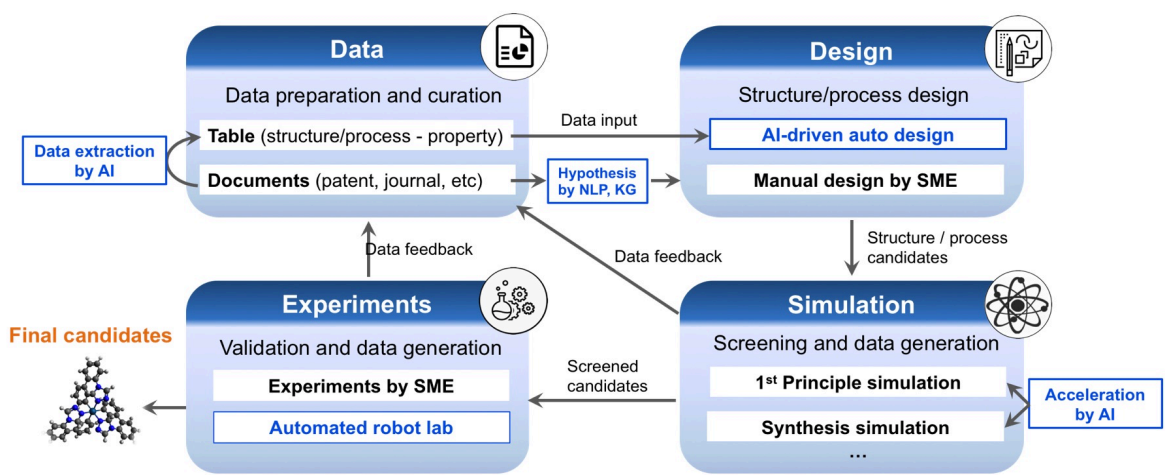

Figure 1. Diagram of general AI-driven material design. AI functions are described by blue-colored characters. This paper focuses on "AI-driven auto design" at the top right.

The ultimate goal of AI-driven automated design is to develop a system that can generate and enumerate material candidate structures that have never existed and that are most likely to possess user-demanded target properties. The methodology to achieve this goal can be divided into two sequential stages: (i) a modeling stage, where the goal is to build an accurate regression or classification model to predict material properties (e.g. glass transition temperature) or attributes (e.g. toxic/non-toxic); and (ii) a design stage, assembling or tuning material structures by directly manipulating molecular structures so that they might possess, based on the trained prediction model, user-demanded target property values (see Figure 2). Related research works, tools, and services in the above stages will be introduced below.

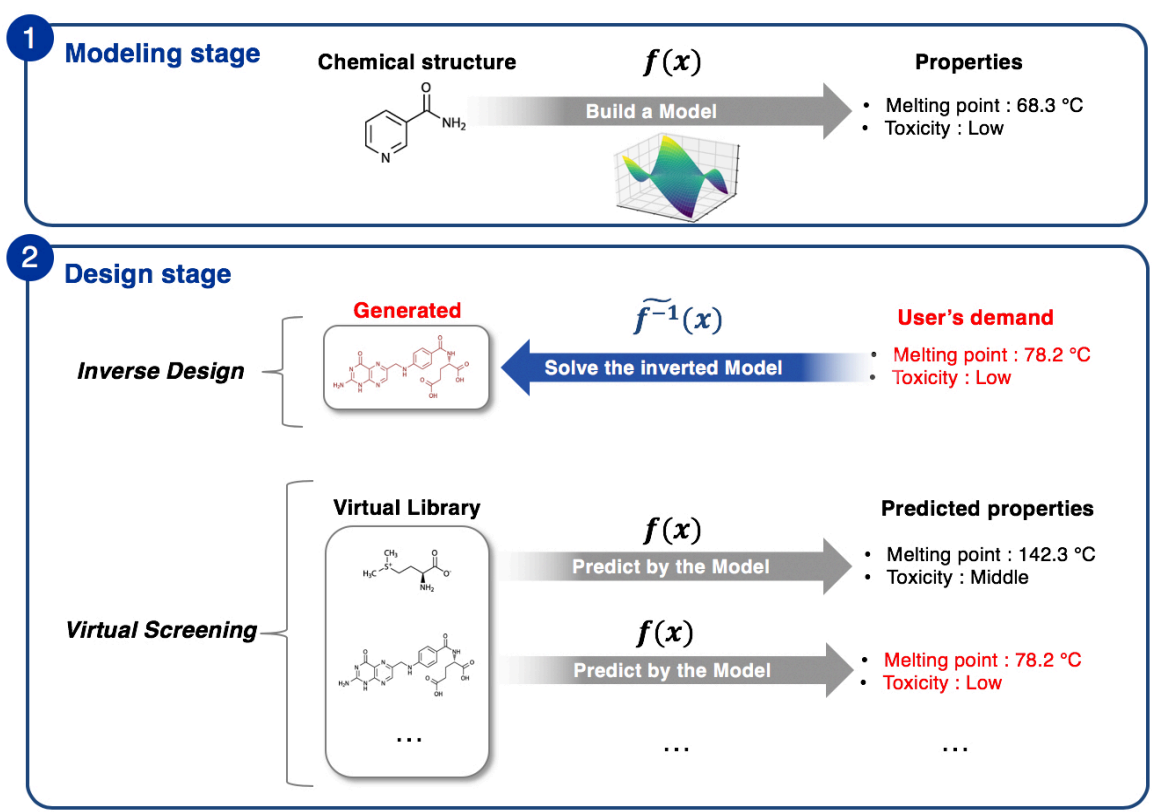

Figure 2. Two stages in material design by AI. Even though not exactly same, those concepts are mostly shared in many methodologies in Cheminformatics.

In the modeling stage, a machine learning or analytics model is trained by a sufficient amount of material data, that is a set of pairs of chemical structure and chemical/physical properties, in order to build a regression or classification model as accurately as possible. In this stage, feature encoding of chemical structures is the key. Especially for organic molecules, various types of feature vector are reported. The most popular feature encoding is the Morgan fingerprint, which captures a molecule's graph topological features by one-hot encoding the appearance of substructures included in a molecule (Rogers et al., 2010).

A more powerful encoding for molecules is Graph Convolutional Neural Network (GCNN) (Faber et al., 2017). There are various versions of GCNN, but their common concept is to iterate summation of neighboring feature vectors defined on each atom in a molecule, then the final feature vector is given by summation (plus some non-linear operation) of all the atom-wise feature vectors (Duvenaud et al., 2015; Gilmer et al., 2017; Altae-Tran et al., 2017). The 3-D molecular structure is well encoded by Coulomb Matrix which encodes inter-atom Coulomb forces in a molecule (Rupp et al., 2012; Montavon et al., 2012; Montavon et al., 2013). For model construction, regression models are mostly built by deep neural networks (DNN) (Pyzer-Knapp et al., 2015), but some classical non-linear models (e.g. Kernel Ridge Regression) also give reasonable prediction results. Most of research work and existing services in this area are focusing on how to increase the accuracy of those prediction models.

In the design stage, material structures are computationally built with reference to a prediction model trained in the modeling stage. Several different approaches exist. The most popular is virtual screening (VS). A large number of chemical structures with unknown properties are thrown into the prediction model, and only the ones predicted to possess desired properties are selected as candidates (Hachmann et al., 2011; Gomez-Bombarelli et al., 2016; Pyzer-Knapp et al., 2015; Pyzer-Knapp et al., 2016). The set of chemical structures with unknown properties is called "virtual library", computationally generated in an exhaustive manner mostly using genetic algorithms that repeat structural manipulation on chemical graphs (e.g., replacing one atom with another, making/breaking a ring, adding/removing a branch) (Virship et al., 2013). The VS approach is useful when users can spare the substantial computational resources required for training models and creating a virtual library large enough (typically more than $10^6$-$10^9$ molecules) to include some promising candidates; but considering its exhaustive search concept, it is not necessarily the best option. The other important approach is inverse design, which solves an inverse problem on the prediction model, and directly generates molecular structures that satisfy user-demanded physical/chemical properties. In contrast to VS, an exhaustive design approach, the inverse design approach, ideally, generates only promising structures. A remarkable example of this approach is SMILES-based generative model leveraging variational autoencoder (VAE) (Gomez-Bombarelli et al., 2017, Guimaraes et al., 2018). SMILES (Simplified Molecular Input Line Entry System) is a form enabling one to describe a chemical structure by using only ASCII strings (Daylight, 1997). The authors constructed a VAE that encodes input SMILES corresponding to one chemical structure to a point in latent space, which is decoded to SMILES identical to the input one. The latent space is connected to another neural network layer having output nodes corresponding to target properties. After training by a dataset (pairs of SMILES and target properties), solution candidates are searched for in the latent space by the descent method and found solution points are decoded to SMILES strings.

Some of technical components included in the above methods are packaged and available as tools or webservices (Deepchem; Polymer Genome; Chainer Chemistry; MOLGEN, etc). Those tools provide very powerful functions for specific purposes (e.g. property prediction, structure generation, etc.), but for the purpose of comprehensive organic material design, technical components should be architected so that they can combine into a coherent workflow from the start: ingest material data relevant to the goal; develop a regression model; and generate material candidates. In order to realize that purpose, we developed a comprehensive inverse design system for small organic molecules, that can provide end-to-end AI-driven material inverse design. In this report we will introduce the fundamental methodology implemented in our system.

## 2. Methods

The fundamental workflow of our approach consists of (0) data input, (1) feature encoding, (2) property prediction, (3) solution search, and (4) structure generation, that are sequentially connected (see Figure 3). First, a data set with table format listing pairs of chemical structures (SMILEs format) and target properties is input as a training dataset. Chemical structures are encoded to a set of feature vectors so that a regression model can be built to predict target properties. This process corresponds to the Modeling stage. Once a model with enough accuracy is built, the Design stage begins. A user inputs a requested property or properties, then the system performs the solution search algorithm to identify feature vector candidates that can satisfy the request. Finally, the feature vector candidates are decoded to concrete chemical structures by the structure generation algorithm. In this section, a detailed expression of each method is described.

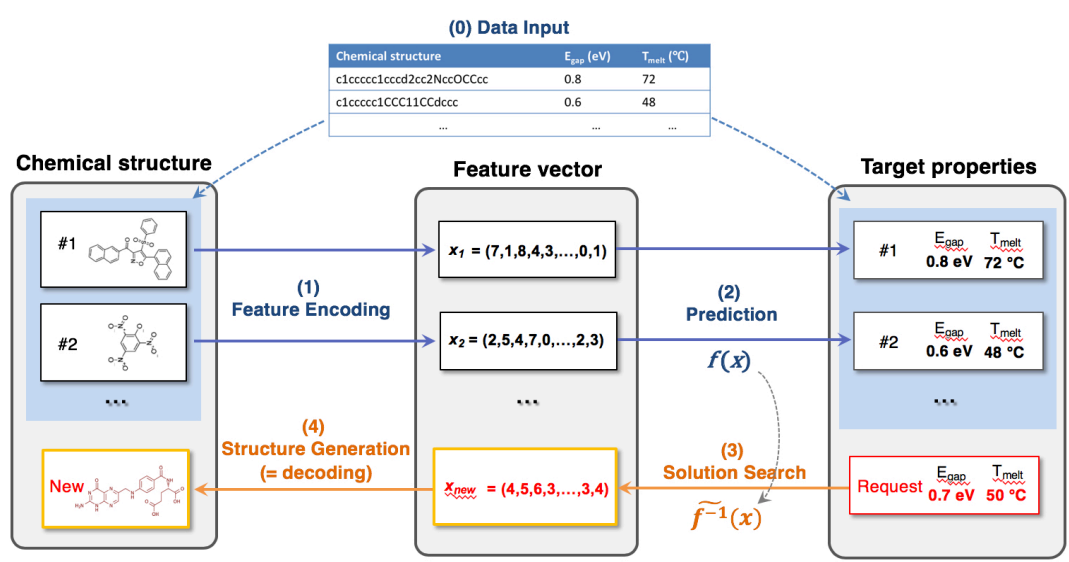

Figure 3. The fundamental workflow of our system. Starting with Data input, technical components are sequentially connected to Structure Generation. The blue-colored and orange-colored arrows correspond to Modeling stage (build property prediction models) and Design stage (inverse design of chemical structures) respectively.

## 2.1 Feature Encoding

Feature vector for chemical structure is required to satisfy four important demands. First, like the feature vector in any industrial domain, it should capture the object's features well so the prediction model can be as accurate as possible. Second, it should not be defined preliminarily by limited human experience or knowledge but instead in an automatic data-driven manner. A frequent request from the industrial side is that they want to explore materials concepts beyond the area governed by their experience-based "common sense". Third, the feature vector should be one from which chemical structures can be generated (corresponding to "Structure Generation" in Figure 3). Finally, it is desirable that the feature vector is interpretable so that users can review, evaluate its adequacy, and adjust it manually if needed.

To satisfy the above requirements, we define substructure-based feature vectors. When we regard a molecule as a graph structure in which atoms and bonds correspond to nodes and edges, a substructure corresponds to a subgraph. The smallest substructure is an atom, and the largest is the molecule itself. The concept of our feature encoding is to count number of substructures included in each molecule and use that array of numbers as a feature vector. First, substructures included in the all molecules in the input dataset are exhaustively identified with limited maximum size (e.g. the maximum number of bonds in each substructure should be three). For each molecule, the identified substructures are counted, and the numbers are arrayed as a vector. An example for two molecules is shown in Figure 4. This approach may be less powerful than GCNN or Coulomb Matrix on the aspect of model accuracy. However, it has some advantages; the most important is that concrete chemical structures can be assembled from the vector values in the "Structure Generation" process, and second, the configured feature vector is completely interpretable.

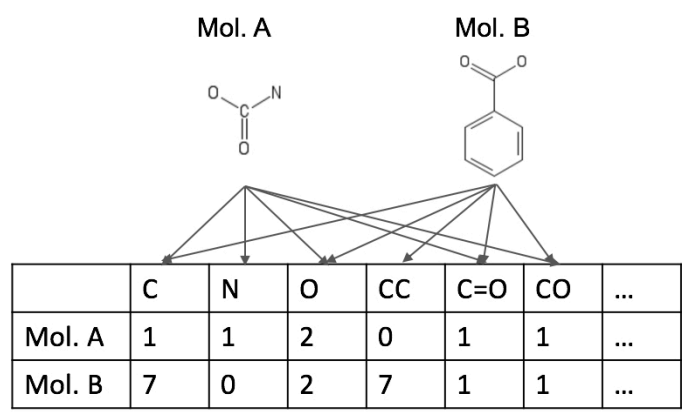

Figure 4. Example to identify and count substructures included in each molecule. Some of substructures are not common for the given molecular set (in this case, "N" and "CC").

The above method is for only a molecular structure, but when a user needs to ingest additional numerical information, simple concatenation works well. In case that the experimental condition also needs to be modeled (e.g. predict a polymer glass transition temperature from monomer structure and polymerization conditions), the quantitative condition values (e.g. heating temperature, injection speed, etc) should be arrayed and concatenated to the molecular feature vector.

**2.2 Property Prediction**

Using the above feature vector, the system builds a regression model to predict target properties. The model should be selected depending on material type and target properties. No particular kind of regression need be used, but common regressions such as kernel ridge regression generally provide adequate accuracy. In case that multiple properties are targeted, models are independently built for each property, but in case of using a DNN, a network with multiple outputs can be an option.

One difference from other machine learning tasks is that a DNN is not necessarily the best option. The reason is that in material industries generally the number of available data is much smaller (typically on the order of $10^1$ to $10^3$) than the case of object recognition, speech recognition, text mining and so on, and therefore the high representative power of a DNN can bring on the problem of over fitting. The main exceptions are where target properties are low-level (i.e. atomistic level properties such as energy bandgap, not mesoscopic molecular level or macroscopic functional level such as luminescent efficiency), so that the properties of a myriad number of molecules can be quickly calculated by physical simulation (e.g., DFT simulation) (Pilania et al., 2013; Mannodi-Kanakkithodi et al., 2016).

**2.3 Solution Search**

After the prediction model is built, a user can set target property values, and the Design stage runs. First, the system generates feature vector candidates that satisfy user-set target properties. Due to non-linearity and complexity of the prediction model, it is difficult or impossible to obtain a direct inverse function of it, therefore the system exploits a search algorithm instead of directly solving an inverse function. The plane of a prediction model has multiple peaks of local minima and its search space is discrete, so gradient methods are not appropriate. We therefore use the Particle Swarm Optimization (PSO) algorithm with a penalty term for molecular constraints as described later. PSO is a powerful and common meta-heuristic optimization algorithm in which virtual "particles" move around to search out solution candidates. Each particle iteratively changes position and velocity to minimize the loss function given by the prediction model.

A crucial point is that the chemically feasible feature vectors distribute on a substantially narrowed manifold that is defined by relationship between the feature vector's elements. Some constraints can be checked explicitly. For example, the number of heavy atoms is limited by the number of substructures (e.g. number of atoms [O] included in a molecule should be greater than or equal to the number of substructures

[-OH] included in the same molecule).

Other constrains are implicit and complicated. The system checks the realizability of a connected graph from the substructures given by a feature vector. In the realizability check, difference of atom symbols and bond types are ignored and only graph structures of the substructures are considered. Degree frequency (the number of vertex of 1, 2, 3, and 4 degree) is counted for the graph structures, and the graph realizability of the degree frequency is checked. Since all the feasible degree frequencies of connected graphs up to given size can be enumerated before the solution search, the system checks if the counted degree frequency reaches one of the feasible degree frequencies with additional vertices. The graph realizability is only a necessary condition of the feasibility, but it is important to screen out some infeasible feature vectors in the solution search considering the huge computational cost of the next phase, Structure Generation.

## 2.4 Structure Generation

The feature vector candidates discovered out in the Solution Search are decoded to concrete molecular structures. For each feature vector, a graph generation algorithm is exploited to assemble molecular structures that satisfy values of feature elements (i.e. the number of substructures, atoms, rings, etc). Molecular structures are generated by connecting atoms and screened by the number of substructures assigned by the feature vector. Of importance is to generate structures both exhaustively (without omission) and efficiently (without isomorphic duplication). A simple process to exhaustively connect atoms leads to divergent combination patterns of atoms, in which most of molecules are isomorphic duplications. In order to make the process efficient, it is important to prune generation paths and avoid the duplications.

We applied McKay's canonical construction path algorithm (McKay, 1998), which is also used in MOLGEN (Kerber al. al., 1998) and OMG (Peironcely et al., 2012), with some modification. McKay's algorithm computes canonical labeling (Hartke and Radcliffe, 2009) for graphs in generation steps to judge whether the graph can be further grown or not from the viewpoint of isomorphism. Graphs are grown by repeating a generation step to add a new vertex to extendable vertices of a current graph (see Figure 5) in a depth first manner starting from an initial single vertex. At each generation step, a canonical labeling of a current graph is computed. The labeling algorithm assigns ordinals to all the vertices uniquely among the isomorphic graphs, which gives a unique construction order of vertex addition for obtaining the graph. If a new vertex to add coincides with the last vertex in the construction order, the generation step continues. Otherwise, the generation step terminates immediately because its construction path generates only a duplicated isomorphic graph.

Graph structures of rings and fused rings are predefined, and all the variations are generated by replacing the vertices with available atoms avoiding isomorphic patterns. They are treated as a single vertex in the generation step. Only when a new vertex is added to them, they are restored to original graph structures. Since the canonical labelling is a bottleneck of the graph generation and its computation time depends on the size of a graph, it greatly improves performance to treat a ring structure as a single vertex when possible.

The numbers of substructures, atoms and rings in the feature elements are updated every time a new vertex is added to the current graph in the generation step. If they satisfy the values of the feature vector, a molecular structure of the graph is stored. The satisfiability of the feature vector in the future generation steps along the current construction path is also checked considering the remaining atoms and the required number of sub-structures, and the generation step terminate if there is no satisfiability.

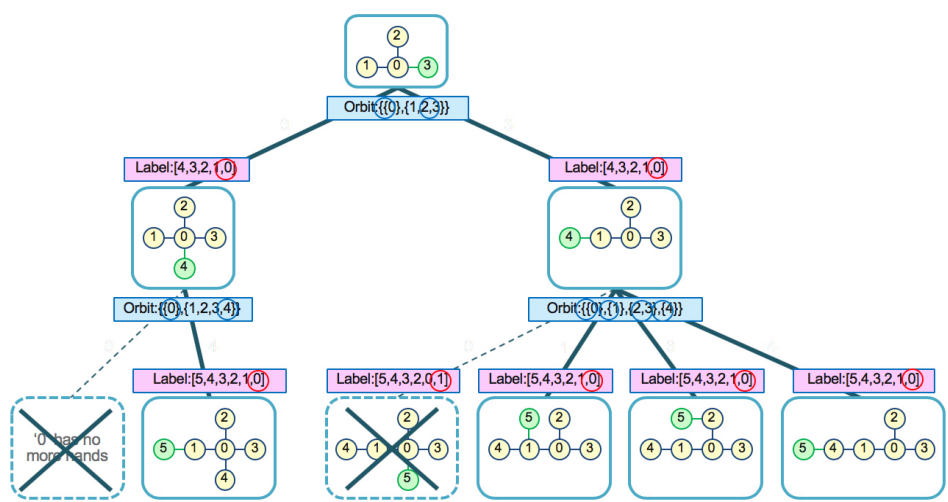

Figure 5. McKay's canonical construction path to generate graph structures. At each time a vertex is added to the current graph, a label is calculated.

## 3. Demonstration

In this section we demonstrate the inverse design of new molecular structures that can satisfy target properties following the steps described above.

### 3.1 Data

For training the system, we used a dataset extracted from the QM9 Dataset (Irwin and Shoichet, 2005; http://zinc15.docking.org/). QM9 is a small organic molecules dataset consisting of 134k molecular structures and associated properties. Each molecular structure is configured by up to 9 heavy atoms with limited atom types (C, O, N, F). Properties include geometric, energetic, electronic, and thermodynamic properties calculated by DFT theory at the B3LYP/6-31G(2df,p) level. For simplicity of demonstration, we use a smaller 1,000 molecule (1k) dataset regularly extracted from QM9, and focus on a physical property; LUMO energy ($\varepsilon_{LUMO}$). The distribution of $\varepsilon_{LUMO}$ of the extracted data is shown in Figure 6. Some of the chemical structures randomly selected from the arrowed area are also shown. A tendency is observed that triangular or squared rings increase, and aromatic rings decrease with increase of $\varepsilon_{LUMO}$ value.

In this demonstration, we try to design new chemical structures that independently satisfy the three ranges; $0.038 \leq \varepsilon_{LUMO} \leq 0.042$ (Ha), $-0.02 \leq \varepsilon_{LUMO} \leq 0.002$ (Ha), $-0.042 \leq \varepsilon_{LUMO} \leq -0.038$ (Ha).

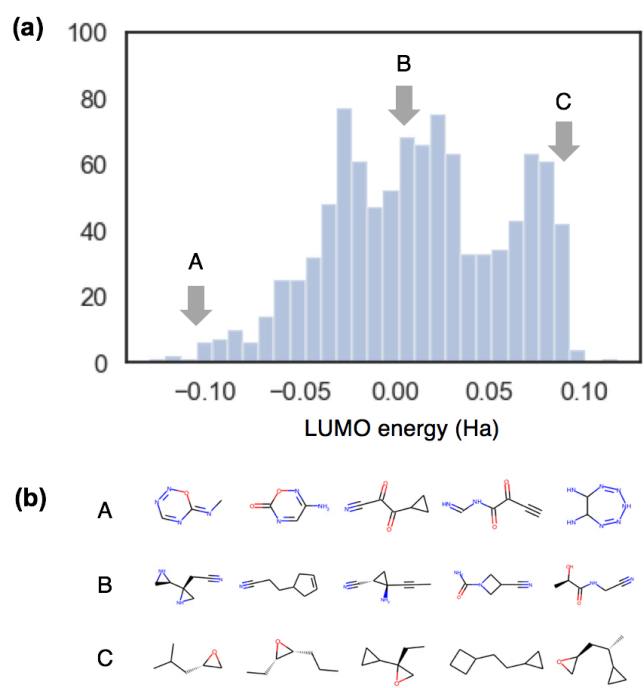

Figure 6. LUMO energy's distribution of the 1,000 (1k) dataset (a). Randomly selected chemical structures of each arrowed area are exhibited in (b).

## 3.2 Feature Encoding

The system encodes molecular structures in the 1k dataset to feature vectors. As described in 2.1, feature vectors are generated by counting number of components included in each molecule. For comparison we generate 6 types of different feature vectors by varying components to count. Common components to count are {heavy atoms, rings, aromatic rings}. Each component includes sub-components, for example {heavy atoms} includes {C, N, O, ⋯}. We call the feature made of this component set "Feature 0". We also add a set of substructures including 1, 2, 3, 4, and 5 bonds to generate Feature 1, Feature 2, Feature 3, Feature 4, Feature 5, respectively (see Figure 7 (a)). Here, it should be noticed that the number of bonds means the number of edges between atoms in graph representation, not a single nor double nor triple bond as in chemistry. Substructures to count are automatically identified referring to the 1k dataset. Concrete components including the substructures are shown in Figure 7 (b). The number of each component corresponds to each element of the feature vectors. As is shown in this figure, with an increase of bond number, substructure patterns get more complex including branches and rings, therefore those feature vectors are expected to have higher representing power. As to aromatic rings, pre-defined aromatic patterns are counted only for 5, 6, and 7 membered rings. Dimensions of feature vectors of Feature 0, Feature 1, Feature 2, Feature 3, Feature 4, and Feature 5 are 13, 32, 97, 380, 1370, and 4356, respectively. Those dimensions will be reduced by feature selection in the regression process.

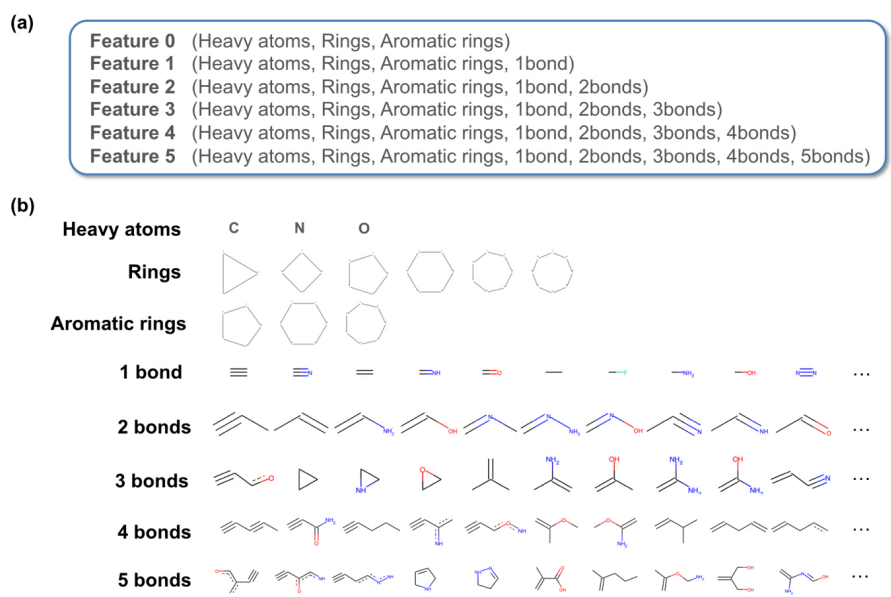

Figure 7. Components included in each feature vector is exhibited in the blue-edged box (a). Subcomponents of each component are exhibited in (b).

### 3.3 Property Prediction

Using the above feature sets, the system builds regression models to predict $\varepsilon_{LUMO}$ by exploiting three regression methods; Lasso, Ridge, and Kernel Ridge with varying hyperparameters. For each regression with a assigned hyperparameter set, the 1k dataset is used for training by 10-fold cross validation, in which the model accuracy is evaluated by $R^2$ (decision coefficient) value. In Lasso regression, the hyper parameter $\alpha$ ($L_1$ penalty term) is swept from $10^{-4}$ to $10^2$ with logarithmic increment. In Ridge regression, the hyper parameter $\beta$ ($L_2$ penalty term) is swept from $10^{-4}$ to $10^2$ with logarithmic increment. In Kernel Ridge regression, we fixed the kernel to RBF (Radial Basis Function) and swept hyper parameters $\alpha$ ($L_2$ penalty term) and $\gamma$ (width of RBF kernel) from $10^{-4}$ to $10^2$ and $10^{-3}$ to $10^2$ respectively with logarithmic increment.

$R^2$ obtained by using the 6 feature sets and the 3 models are plotted in Figure 8. Each $R^2$ is an average value on the 10-fold cross validation. As is shown, in any model there is a tendency that the model accuracy increases as the feature set gets plentiful, but it saturates when the feature set includes substructures having more than 2 bonds. From the viewpoints of accuracy and feature vector dimension, we select Feature 2 (97-dimension) and the Kernel Ridge regression model as the final prediction model. The regression results of this model are shown in Figure 9.

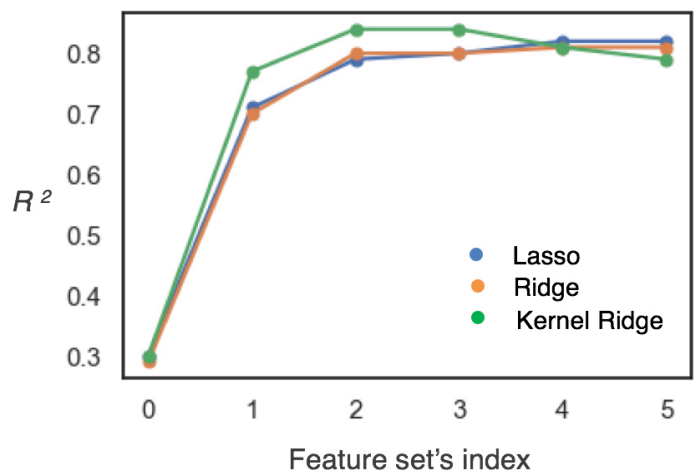

Figure 8. Accuracy of prediction model as a function of Feature set's index defined in Figure 7, and three types of regression models.

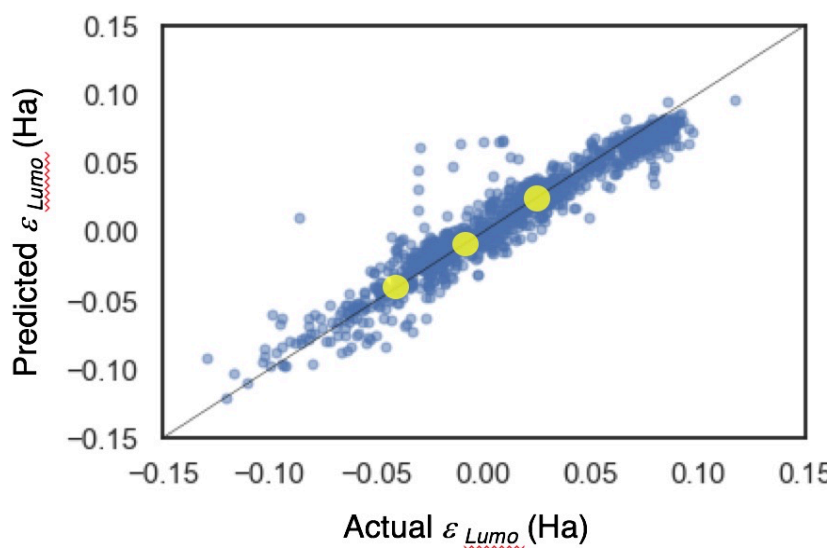

Figure 9. Regression result to predict LUMO energy. Kernel Ridge regression model with Feature 2 was used. Yellow circles represent the targeted area for inverse design process.

3.4 Solution Search

By using the prediction model, the system runs the solution search algorithm, PSO with chemical constraint, to search out feature vector candidates satisfying the three target value ranges. The search space (range of feature vector's elements) is automatically set considering the original data's distribution. After running the PSO using 1000 particles, we obtained 30 feature vectors (the number to obtain can be set arbitrarily) for each target $\varepsilon_{LUMO}$ value range. To compare the feature distributions, we performed PCA (Primary Component Analysis) on feature vectors, and plotted them on two-dimensional space (see Figure 10). As shown, it is observed that feature vectors searched out for the three $\varepsilon_{LUMO}$ values ranges are distributing with some degree of divergent profile. This confirms that the PSO algorithm successfully searched out solutions in divergent areas.

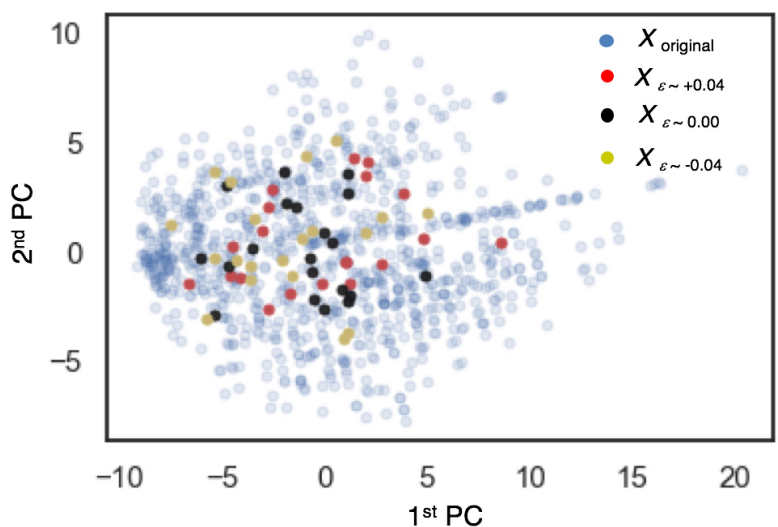

Figure 10. Distribution profile of 1st and 2nd principle components of feature vectors. Blue plots represent the original dataset. Red, black, and yellow plots represent searched out feature vectors for each LUMO energy.

### 3.5 Structure Generation

Finally, chemical structures are automatically generated from each feature vector candidate by running the generation algorithm described in 2.4. Despite constraining the search space by a penalty term in PSO, some of the feature vector candidates still include conflicting feature element values. Generation process using such an infeasible feature vector fails. The generation success rate ranges from 10-50%, depending on target property's type and values, model accuracy, etc. For the three ranges $0.038 \leq \varepsilon_{LUMO} \leq 0.042$ (Ha), $-0.02 \leq \varepsilon_{LUMO} \leq 0.002$ (Ha), $-0.042 \leq \varepsilon_{LUMO} \leq -0.038$ (Ha), we obtained 4, 6, and 5 feasible feature vectors out of 30, respectively. From each feasible feature vector, 1 to 10 chemical structures are generated. Part of the generated structures are shown in Figure 11. Searching the original 1k dataset, it is confirmed that they are not included in the dataset, which means a set of "brand new" chemical structures were automatically invserse-designed by the system.

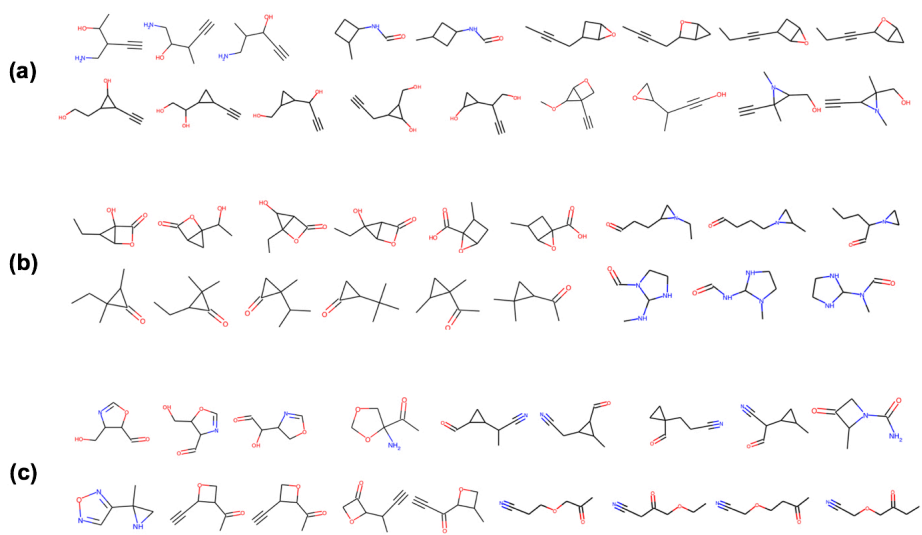

Figure 11. Example of the automatically generated chemical structures satisfying the three targeted LUMO energy values; (a) $0.038 \leq \varepsilon_{LUMO} \leq 0.042$ (Ha), (b) $-0.02 \leq \varepsilon_{LUMO} \leq 0.002$ (Ha), and (c) $-0.042 \leq \varepsilon_{LUMO} \leq -0.038$ (Ha).

## 4. Limitation and Customization

The methodology introduced in this paper supports only the most fundamental algorithms that are common for organic material applications. In order to make the system more powerful for each material domain, user-defined domain specific customizations will be required. For example, the simplest is to set structural rules in the generation algorithm (e.g. forbidden to include some user-defined substructures). Another example is to concatenate user-defined feature vectors (e.g. polymerization conditions) to the automatically generated feature vectors. Further advanced customization calls for functional level improvement; generate feature vectors for 3-dimensional conformation, etc. Those advanced customizations can be achieved by collaboration with domain-experts. More complex features may also be required to represent and sample other classes of materials, such as inorganic crystals. Also, in this approach synthetic accessibility or chemical relevance of structures are not considered. If SME can score structures for synthetic accessibility, and set those values as corresponding target properties, they can be considered in the modeling stage.

## 5. Tool in Service

The methodology and algorithms described in this paper are implemented as a comprehensive material design tool in the format of a Python package. Some of the low-level functions are developed by open source packages: RDKit (https://www.rdkit.org/) for manipulating and drawing chemical structures, Scikit-learn

([https://scikit-learn.org/stable/](https://scikit-learn.org/stable/)) for common regression models. Each of the building blocks (e.g. feature encoder) in the workflow are treated to be single or multiple microservices which are composed as Python modules. These modules are communicated and integrated automatically to provide an end-to-end material design service. The tool is deployed on the cloud and is accessible for our client users. By harnessing the power of scalable cloud resources and services, our material design platform eliminates the hurdles that accompany traditional material discovery tools: hardware requirements, installation troubles, version compatibility, and required feature/security updates. Today the tool is powered by JupyterHub ([https://jupyterhub.readthedocs.io/en/stable/#](https://jupyterhub.readthedocs.io/en/stable/#)) which enables worldwide lab members to create individual workspaces via a generally used web browser. Documentations in HTML format narrative text and live code bases greatly improve information exchange and sharing.

In the near future, the platform will be expanded by implementing a full Web application with interactive console and REST API where the modules described above are combined into a framework which automates common tasks and where interactive documentation are generated by Swagger UI.

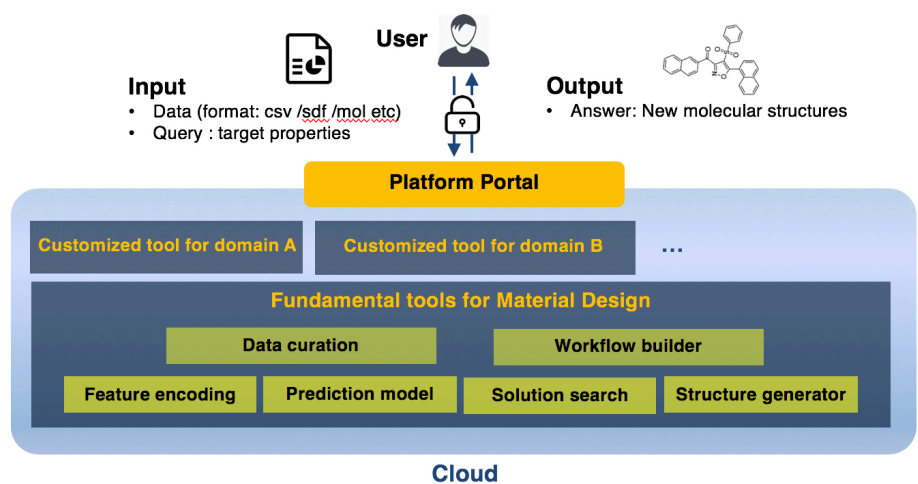

Figure 12. Architecture of our system. Functional components are provided as Python modules (yellow boxes). Those modules are automatically orchestrated by workflow building modules. This tool set works by itself as demonstrated in this paper, but additional customization enhances the functions for specific material domains.

6. Summary

In this paper, we introduced the fundamental methodology of our automated material design system. In this methodology, technical components are architected to be sequentially integrated so that they comprise an end-to-end workflow; feature encoding, property prediction, solution search, and structure generation. We also demonstrated to inverse-design new molecular structures that satisfy the targeted LUMO energies.

The system is deployed on cloud and provided to our client users across broad material domains; engineering polymer, drug, color material, etc.

## Acknowledgments

This work was conducted under IBM Research Frontiers Institute with its member companies; Canon, Hitachi Metals, Honda, JSR, Nagase, and Samsung (Alphabetical order). We thank them for their collaboration and feedback about our system.

## Reference

R. S. Bohacek, C. McMartin, and W. C. Guida, The art and practice of structure-based drug design: A molecular modeling perspective, Medicinal Research Reviews, 16 (1), 3-50 (1996).
Byung-Kwon Choi et al., Literature-based automated discovery of tumor suppressor p53 phosphorylation and inhibition by NEK2 (2018)
Haichen Li, et al., Tuning the Molecular Weight Distribution from Atom Transfer Radical Polymerization Using Deep Reinforcement Learning (2018)

D. Rogers and M. Hahn, Extended-connectivity fingerprints, Journal of chemical information and modeling, 50(5), 742-754 (2010).

F. A. Faber et al., Machine learning prediction errors better than DFT accuracy, arXiv:1702.05532v2 (2017)
D. Duvenaud, Convolutional networks on graphs for learning molecular fingerprints, Proceedings of the 28th International Conference on Neural Information Processing Systems, 2224-2232 (2015).
J. Gilmer, et al., Neural message passing for quantum chemistry, arXiv:1704.01212v2 (2017)
H. Altae-Tran, et al., Low data drug discovery with one-shot learning, ACS Central Science 3, 283 (2017)

M. Rupp et al., Fast and accurate modeling of molecular atomization energies with machine learning, Phys. Rev. Lett., 2018, 053801 (2012).

Montavon et al., Learning invariant representations of molecules for atomization energy prediction, 26th Annurual Conference on Neural Information Processing Systems 2012 (NIPS 2012), Lake Tahoe, NV, US, 440-448.

Montavon et al., Machine learning of molecular electronic properties in chemical compound space, New Journal of Physics, 15 (2013).
E. O. Pyzer-Knapp, K. Li, and A. A-Guzik, Learning from the Harvard Clean Energy Project: The Use of Neural Networks to Accelerate Materials Discovery, *Materials Review*, 25, 6495-6502 (2015).
J. Hachmann et al., The Harvard Clean Energy Project: Large-scale computational screening and design of organic photovoltaics on the world community grid, J. Phys. Chem. Lett., 2 (17), 2241-2251 (2011).
R. Gomez-Bombarelli et al., Design of efficient molecular organic light-emitting diodes by a high-

throughput virtual screening and experimental approach, Nature Materials, 15, 1120-1128 (2016).

E. O. Pyzer-Knapp, G. N. Simm, and A. A-Guzik, A Bayesian approach to calibrating high-throughput virtual screening results and application to organic photovoltaic materials, *Mater. Horiz.* (2016).

E. O. Pyzre-knapp et al., What is high-thoughput virtual screening? A perspective from organic materials discovery, Annual Review of Materials Research, 45, 195-216 (2015).

A. M. Virship et al., Stochastic Voyages into Uncharted Chemical Space Produce a Representative Library of All Possible Drug-Like Compounds, *Journal of the American Chemical Society*, 135, 7296-7303 (2013).

R. Gomez-Bombarelli et al., Automatic chemical design using variational autoencoders, arXiv:1610.02415v3.

G. L. Guimaraes, Objective-reinforced generative adversarial networks (ORGAN) for sequence generation models, [arXiv:1705.10843v3](arXiv:1705.10843v3) (2018)

Daylight Chemical Information Systems, Inc., SMILES – A simplified chemical lnguage. http://www.daylight.com/dayhtml/doc/theory/theory.smiles.html (accessed 9-Feb-2018)

DeepChem, https://deepchem.io/

Polymer Genome https://www.polymergenome.org/

Chainer Chemistry, https://github.com/pfnet-research/chainer-chemistry

MOLGEN, http://www.molgen.de/

G. Pilania et al. Accelerating materials property predictions using machine learning, *Scientific Reports*, 3, 2810 (2013).

A. Mannodi-Kanakkithodi et al., Machine learning strategy for accelerated design of polymer dielectrics, Scientific Reports, 6, 20952 (2016).

B. D. McKay. Isomorph-free exhaustive generation. *J. Algorithms*, 26(2):306–324, 1998

A. Kerber, R. Laue, T. Grüner, M. Meringer, MOLGEN 4.0, *MATCH Communications in Mathematical and in Computer Chemistry 37*, 205-208, 1998

S. G. Hartke and A. Radcliffe. Mckay's canonical graph labeling algorithm. In *Communicating Mathematics*, 479, 99–111. American Mathematical Society, 2009.

J. E. Peironcely et al, OMG: Open Molecule Generator, *Jouranal of Cheminfomatics* 4:21, 2012

J. J. Irwin and B. K. Shoichet, ZINC – A free database of commercially available compounds for virtual screening, J. Chem. Inf. Model. 45 (1), 177-182 (2005).